\documentclass[
reprint,
superscriptaddress,
 amsmath,amssymb,
 aps,
]{revtex4-2}

\usepackage{graphicx}
\usepackage{hyperref}

\hypersetup{
    colorlinks = true,
    linkcolor= blue,
    citecolor= blue,
    urlcolor= blue
}

\usepackage{amsmath}
\usepackage{xcolor}
\usepackage{bm}
\usepackage{amsmath}

\begin{document}

\title{Orientational ordering in active nematic solids}

\author{Haiqian Yang}
\affiliation{Department of Mechanical Engineering, Massachusetts Institute of Technology, Cambridge, Massachusetts 02139, USA}
\author{Ming Guo}
\affiliation{Department of Mechanical Engineering, Massachusetts Institute of Technology, Cambridge, Massachusetts 02139, USA}
\author{L. Mahadevan}\thanks{lmahadev@g.harvard.edu}
\affiliation{School of Engineering and Applied Sciences, Harvard University, Cambridge, Massachusetts 02138, USA}
\affiliation{Department of Organismic and Evolutionary Biology, Harvard University, Cambridge, Massachusetts 02138, USA}
\affiliation{Department of Physics, Harvard University, Cambridge, Massachusetts 02138, USA}

\date{\today}

\begin{abstract}

\textit{In vivo} and \textit{in vitro} systems of cells and extra-cellular matrix (ECM) systems are well known to form ordered patterns of orientationally aligned fibers. Here, we interpret them as active analogs of the (disordered) isotropic to the (ordered) nematic phase transition seen in passive liquid crystalline elastomers. A minimal theoretical framework that couples cellular activity (embodied as mechanical stress) and the finite deformation elasticity of liquid crystal elastomers sets the stage to explain these patterns.  Linear stability analysis of the governing equations about simple homogeneous isotropic base states shows how the onset of periodic morphologies depends on the activity, elasticity, and applied strain, provides an expression for the wavelength of the instability, and is qualitatively consistent with observations of cell-ECM experiments. Finite element simulations of the nonlinear problem corroborate the results of linear analysis.  These results provide quantitative insights into the onset and evolution of nematic order in cell-matrix composites.

\vspace{20pt}
\end{abstract}

\newcommand{\mat}{\text{\tiny R}}
\newcommand{\snA}{{\mathsf{A}}}
\newcommand{\trans}{{\mskip-2mu\scriptscriptstyle\top}} 
\newcommand{\tendot}{\mskip-3mu:\mskip-2mu}

\def\I{\mathbf{I}}
\def\Q{\mathbf{Q}}

\def\X{\mathbf{X}}
\def\x{\mathbf{x}}
\def\F{\mathbf{F}}
\def\FT{\mathbf{F^\trans}}
\def\C{\mathbf{C}}
\def\E{\mathbf{E}}

\def\e{\bm{\varepsilon}}
\def\edev{\bar{\e}}

\def\S{\bm{\sigma}}

\def\T{\mathbf{T}}
\def\P{\mathbf{P}}

\def\n{\mathbf{n}}
\def\t{\mathbf{t}}
\def\v{\mathbf{v}}
\def\u{\mathbf{u}}

\maketitle

Nematic patterns formed by cell-matrix mixtures are ubiquitous within animal organs and tissues~\cite{neville1993biology}. \textit{In vivo}, these network structures spontaneously form due to the interaction between the active cells and the passive extra-cellular matrix (ECM). \textit{In vitro}, many cell-matrix systems can also spontaneously break symmetry and develop nematic network-like patterns~\cite{vader2009strain,weidenhamer2013influence,sander2011initial,doha2022disorder}. However,  a quantitative framework for how cellular activity couples with the passive matrices and explains the transition between disordered isotropic states and ordered nematic states remains an open question. 

While the study of active nematic fluids is a well-trodden area~\cite{doostmohammadi2018active},  patterning driven by activity in nematic solids is much less studied. Studies on cell-matrix interaction have been mainly in the single-cell setting focusing on modeling fiber buckling under compression~\cite{han2018cell,hall2016fibrous}. Effective theories that couple activity to the dynamics of isotropic viscoelastic solids show that cell contraction can induce cell density separation in an active solid~\cite{weber2018differential,yin2023contractility}, but do not account for the role of orientational ordering.  Here we attempt to remedy this by constructing a continuum-mechanical theoretical framework that couples cellular activity with nematic order and elasticity in the context of the onset and evolution of nematic patterns from a homogeneous and isotropic cell-matrix mixture.

\begin{figure}[h!]
    \centering
    \includegraphics[width=0.48\textwidth]{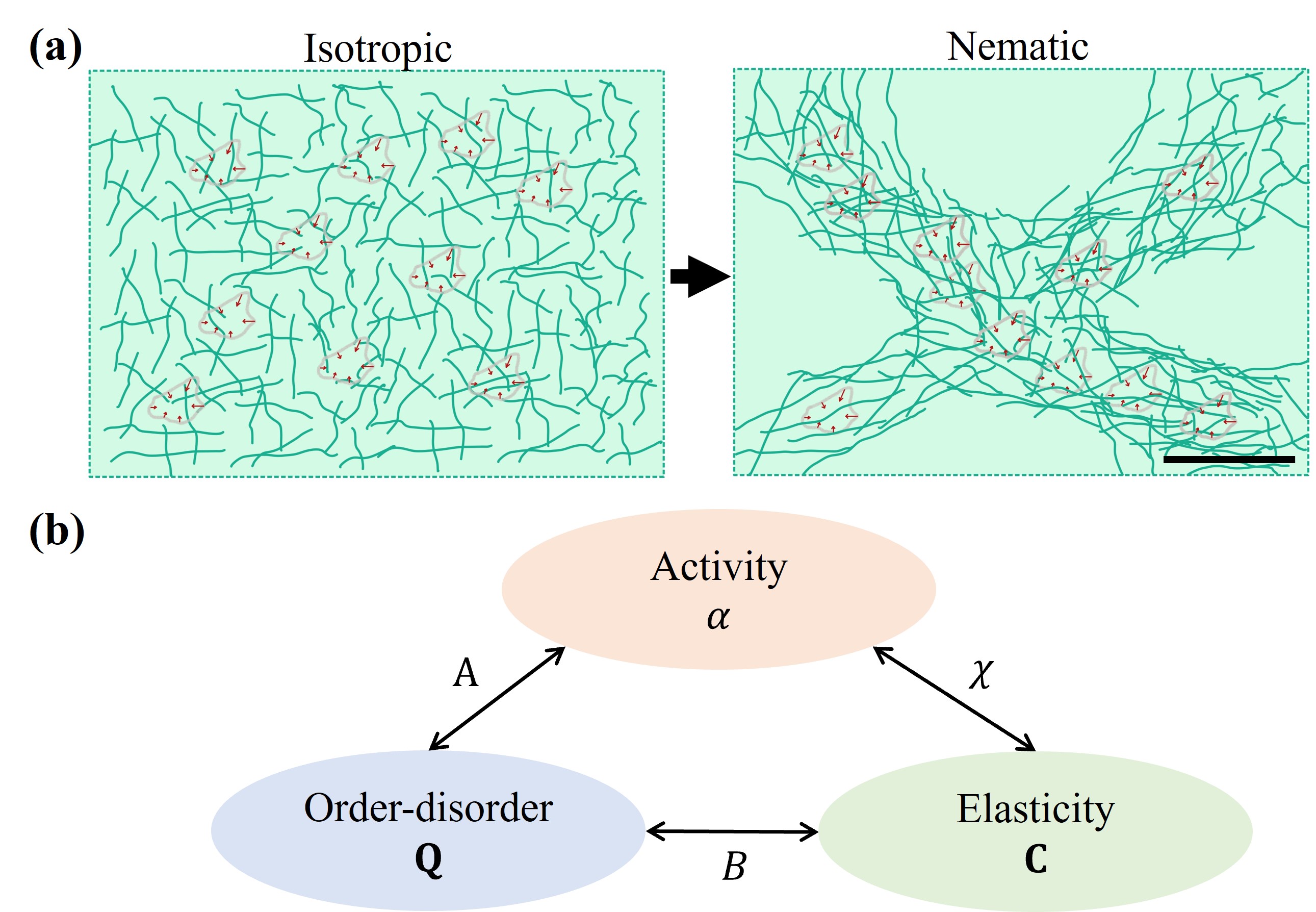}
    \caption{
    (a) A schematic figure showing cell-matrix systems exhibiting a transition from isotropic phase to nematic phase. The cell boundaries are drawn in gray with red arrows indicating cell contraction. The matrix is drawn in green. 
    Scale bar, 50 $\mathrm{\mu m}$.
    (b) The three major aspects of our theory are elasticity, as quantified by the deformation tensor $\C$ (Eq.~\ref{eq:forceBalance}), order-disorder transition, as quantified by the nematicity tensor $\Q$ (Eq.~\ref{eq:evolutionQ}), and activity given by $\alpha$ (Eq.~\ref{eq:evolutionA}).}
    \label{fig:1}
\end{figure}

In Fig.~\ref{fig:1}(a), we schematize how cell activity, matrix mechanics, and the nematic-isotropic transition can lead to spontaneous phase separation and orientational ordering. This serves as the inspiration for a theory of \textit{``active nematic solids"} that involves coupling cellular activity with the classical Landau-de Gennes liquid-crystal theory in a nonlinearly elastic background matrix [Fig.~\ref{fig:1}(b)], where active stress, an embodiment of the cellular activity,  drives the active cell-matrix composites towards an ordered state.

%
\bigskip
\textit{Kinematics.}---
We start by considering a body $\mathbf{\Omega}$ with the region of space it occupies in a fixed reference configuration and denote by $\mathbf{X}$ an arbitrary material point of $\mathbf{\Omega}$. A deformation $\boldsymbol{\phi}: \mathbf{\Omega} \rightarrow \bm{\mathit{\Omega}}$ is a smooth one-to-one mapping $\mathbf{x} = \boldsymbol{\phi}(\mathbf{X}, t)$ with deformation gradient given by $\F = \nabla \boldsymbol{\phi}$,  with $\nabla \equiv \frac{\partial}{\partial \X}$ denoting derivatives with respect to the reference configuration.
Then, the displacement gradient tensor is $\nabla \u = \F-\I$, the volumetric deformation is $J=\text{det}\, \F$, the Cauchy-Green deformation tensor is $\C = \F^\trans \F$, and the distortional deformation tensor is $\bar{\C} = J^{-1} \C$. In terms of the distortional deformation tensor, the first invariant is $\bar{I}_1 = \text{tr}\, \bar{\C}$, and the deviatoric strain is $\bar{\E} = \frac{1}{2} \left(\bar{\C}-\I\right)$. The material is also characterized through the traceless symmetric nematic tensor $\Q (\mathbf{X}, t)$, which can be decomposed as $\Q = S(2\n \otimes \n - \I)$, where $S$ is the nematic order parameter, and $\n = \left( \cos{\theta}, \sin{\theta} \right)^\trans$ is the director. Though here cell density is not explicitly considered, $S$ is a coarse-grained measure that is a function of both the local microscopic alignment and cell density.

\textit{Free energy.}---
Considering a $L \times L$ 2-D periodic domain, the dimensionless coordinate is given by $\X \rightarrow \X/L$. To account for cell activity, matrix elasticity, and the transition between ordered aligned states and disordered isotropic states, we assume that the dimensionless free energy per unit reference volume takes the following form
\begin{equation}\label{eq:energy}
\begin{aligned}
    \Psi_\mat = \frac{\tilde{\Psi}_\mat}{\tilde{G}} = 
     &\underbrace{\frac{1}{2} (\snA+\snA_0) \Q:\Q + \frac{1}{4} C (\Q:\Q)^2}_{\text{Landau-de Gennes}} 
    +\underbrace{\frac{1}{2}K |\nabla \Q|^2}_{\text{Frank}}\\
    +&\underbrace{\frac{1}{2} (\bar{I}_1 -2) + \frac{1}{2}\kappa (J-1)^2}_{\text{neo-Hookean}}\\
    -&\underbrace{B \bar{\E}:\Q - B'(J-1)\Q:\Q}_{\text{Nematoelastic}},\\
\end{aligned}
\end{equation}
where the different terms correspond to contributions associated with nematic order, gradients in nematic order, elasticity and the coupling between these fields; the nematoelastic coupling includes the leading-order deviatoric and volumetric terms as described in~\cite{lubensky2002symmetries}. Here $\tilde{G}>0$ is the shear modulus, and the dimensionless parameters $C>0$, $B>0$, $\kappa>0$, $\snA_0>0$, $K>0$. Here the activity $\snA$ is assumed to be a spatio-temporal dynamic variable that in passive liquid crystals will be a function of temperature, but here will be assumed to be a function of activity. 
In the absence of activity, the terms in the free energy are commonly found in the liquid crystal (elastomer) literature~\cite{warner2007,duzgun2015dynamic,biggins2009supersoft}.
We note that $\Psi_\mat$ approaches classical models in two limits: when $\Q= 0$, $\Psi_\mat$ describes a passive elastomer, and when $\F=\I$, $\Psi_\mat$ describes a passive liquid crystal. The variable $\snA$ serves to control nematic order via active elastic stresses (instead of temperature in the passive case), and further allowed to evolve according to an equation to be introduced later. If $\snA+\snA_0>0$, the Landau-de Gennes energy has a single well, and $\Q = 0$ (isotropic) is energetically favorable; if $\snA+\snA_0<0$, the Landau-de Gennes energy is double-welled and $\Q \neq 0$ (nematic) is energetically favorable.

\begin{figure}[h!]
    \centering
    \includegraphics[width=0.48\textwidth]{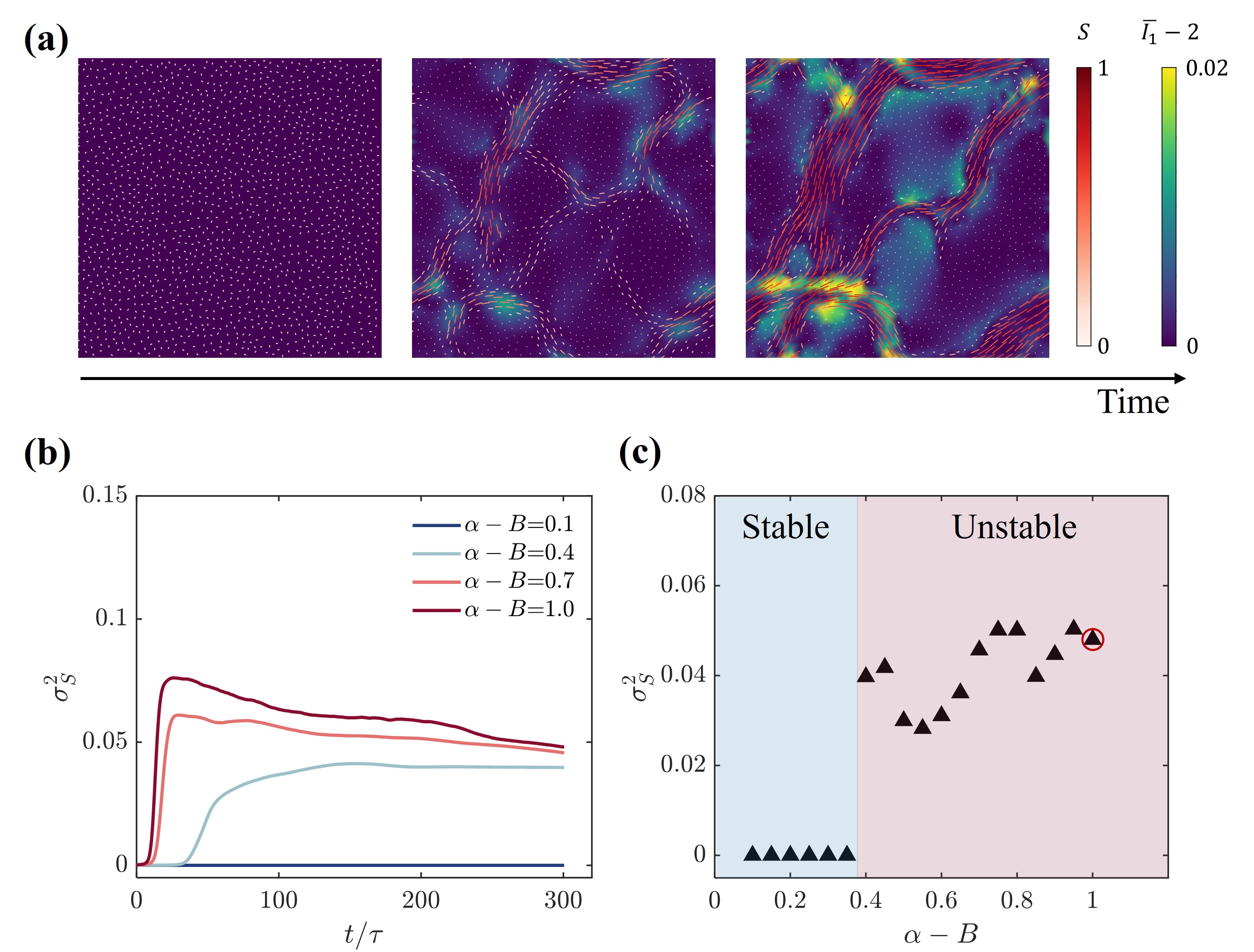}
    \caption{Activity induces spontaneous symmetry breaking and network patterning. 
    (a) Representative snapshots of the simulation. 
    The quivers indicate the director fields. 
    The color of the quiver indicates the magnitude of the nematic order parameter $S$, and the color of the background indicates the magnitude of the deformation $\bar{I}_1 = \text{tr}\, \bar{\C}$.
    (b) The variance of nematic order parameter $\sigma^2_S$ is plotted as a function of time at different cell contractility $\alpha - B$. 
    (c) The variance of nematic order parameter $\sigma^2_S$ at $\frac{t}{\tau}=300$ is plotted as a function of the cell contractility $\alpha - B$. The red circle indicates the point corresponding to panel (a).
    The results here are produced by numerically solving Eqs. (\ref{eq:forceBalance},~\ref{eq:evolutionQ},~\ref{eq:evolutionA}), with $S_0=0.1, \snA_0 = 0.1, \frac{\gamma}{\tau}=1, B = 0.5, K=5 \times 10^{-5}, \chi=10$.}
    \label{fig:2}
\end{figure}

\textit{Force balance.}---
From the point of view of elasticity, it is useful to consider the dimensionless second Piola stress
\begin{equation}\label{eq:stress}
\mathbf{S} = 2\frac{\partial \Psi_\mat}{\partial \C}
    + \underbrace{\alpha \Q}_{\text{Activity}},
\end{equation}
where $\alpha \Q$ is the active stress generated by polarized cells, with $\alpha>0$ the magnitude of active contraction, and $2\frac{\partial \Psi_\mat}{\partial \C}=-B\Q+\bar{\E}+\kappa(J-1)\I - B'(\Q:\Q) \, \I$. We note that $\alpha-B > 0$ in this study, meaning that cells are contractile, and they generate tensile stress in the matrix. Local force balance in the reference configuration implies that
\begin{equation}\label{eq:forceBalance}
\begin{aligned}
           & \text{Div}\, \P = \mathbf{0},\\
\end{aligned}
\end{equation}
where the first Piola stress $\P$ is given by $\P = \F \mathbf{S}$.

\begin{figure}[h]
    \centering
    \includegraphics[width=0.42\textwidth]{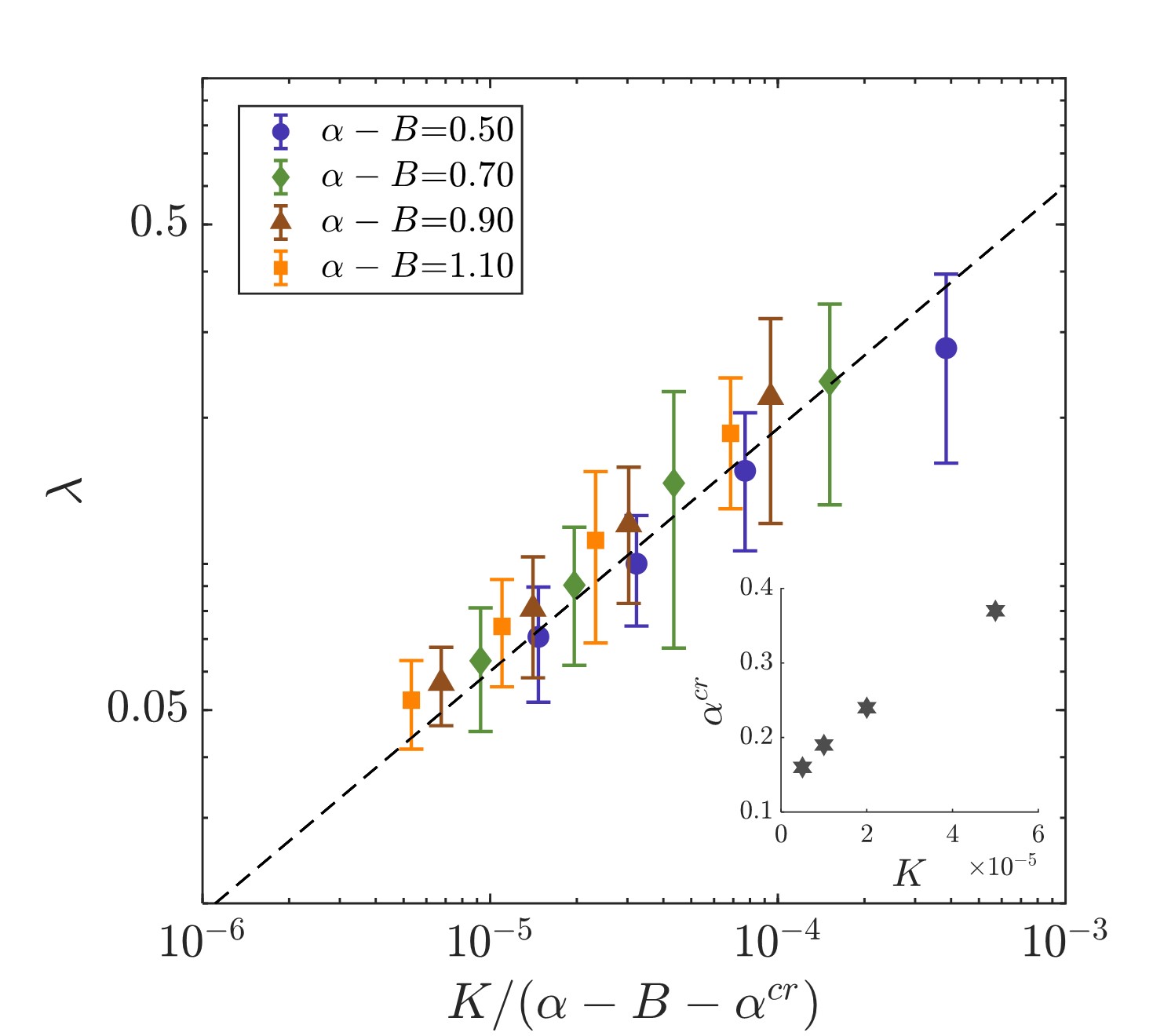}
    \caption{
    Pattern size $\lambda$ vs normalized effective activity $K/(\alpha-B-\alpha^{cr})$. 
    $\lambda$ is calculated using the system configuration corresponding to the first peak value of $\sigma^2_S$ over time. 
    The dashed reference line has a slope of $1/2$.
    The inset panel shows the critical activity $\alpha^{cr}$ as a function of the Frank constant $K$.
    Here $S_0=0.1, \snA_0 = 0.1, \frac{\gamma}{\tau}=1, B = 0.5, \chi =10$.
    The numerical results are from solving Eqs. (\ref{eq:forceBalance},~\ref{eq:evolutionQ},~\ref{eq:evolutionA}).
    Each condition is simulated with 50 different initial random conditions. The marker indicates the mean and the error bar indicates the standard deviation.
    }
    \label{fig:3}
\end{figure}

\textit{The dynamics of nematic ordering.}---
Turning now to how the nematic tensor evolves, we write 
~\footnote{
Effectively, the presence of activity offsets the nematic stress, making the contribution of nematic-elastic coupling in Eq.~\ref{eq:evolutionQ} and in Eq.~\ref{eq:stress} not conjugate. For contractile cells considered in this study, we note that $\alpha - B > 0$.}
\begin{equation}\label{eq:evolutionQ}
    \gamma \mathring{\Q} = -\frac{\delta \Psi_\mat}{\delta \Q} \, ,
\end{equation}
where $\mathring{\Q}=\dot{\Q}-\mathbf{W}\Q+\Q\mathbf{W}$ denotes the Jaumann derivative (consistent with a corotational derivative for a 2-tensor), with the spin tensor $\mathbf{W}=\frac{1}{2} \left( \nabla \dot{\u} - \nabla \dot{\u}^{\trans} \right)$~\cite{sonnet2012dissipative,wang2022nonlinear}, and 
$\frac{\delta \Psi_\mat}{\delta \Q} = \frac{\partial \Psi_\mat}{\partial \Q} + K\nabla^2\Q
= - (\snA+\snA_0+\frac{1}{2}CS^2)\Q+B\bar{\E}+K\nabla^2\Q$. 
Here, the characteristic time scale of $\Q$ is given by $\frac{\gamma}{\snA_0}$.

\textit{Stress-driven orientational ordering.}---
We finally turn to account for how extra-cellular-matrix (ECM) fibers tend to align in the tensile direction, even as cells prefer to line up along the fibers. To account for this active behavior of the composite of cells+ECM, we allow $\snA$ to be a spatially varying variable, inspired by the classical Landau-de Gennes theory of phase transition where $\snA$ is controlled by temperature. Here we assume a phenomenological relation where stress acts as an effective temperature that drives orientationally ordering phase, so that the variable $\snA$ evolves according to
\begin{equation}\label{eq:evolutionA}
    \tau \dot{\snA} = -\zeta \snA - f(\P, \nabla\P) - s \snA^3,
\end{equation}
where the coefficients $\tau>0$, $\zeta >0$ and $s>0$. This minimal model allows the coefficient of the quadratic term in the free energy $\Psi_\mat$ given by $\snA+\snA_0$ to switch sign when driven by stresses, consistent with observations; in the absence of any stress, this coefficient relaxes to its initial value of $\snA_0$.  We assume that $f(\cdot,\cdot)$ depends on the first Piola stress $\P$ because $\snA$ is defined in the reference configuration (we also considered a dependence on the second Piola stress $\mathbf{S}$, but this only affects our results quantitatively).
We assume that activity responds faster than the relaxation of nematic order, and therefore $\frac{\tau}{\zeta} \ll \frac{\gamma}{\snA_0}$. 
Further, while the activity-stress coupling function $f(\P,\nabla \P)$ can be an arbitrary function of the stress and its gradient, here we only consider the first invariant of $\P$ for simplicity; that is $ f(\P,\nabla \P) = \chi \mathrm{tr}\,\P$, with the coefficient $\chi>0$, meaning that tension promotes activity, while compression reduces activity.

\textit{Activity induces spontaneous symmetry breaking.}---
To gain some analytical insight into the coupling between deformation and order-disorder transition, we reduce our finite-deformation theory to incompressible linear elasticity and consider a simplified 2-D problem with vanishing initial nematic order $S_0=0$, but non-zero deformation $\bm{\epsilon} \neq 0$ applied through the boundary. Specifically, we consider a 2-D periodic domain subjected to a uniform shear deformation with components $\epsilon_{11} = \epsilon,\, \epsilon_{22} = -\epsilon,\,\epsilon_{12} = \epsilon_{21} = 0$, with spatially homogeneous initial conditions $\u = 0, \snA = 0$, and vanishing initial nematic order $S_0=0$. We perform linear stability analysis by assuming a small perturbation $\propto \mathrm{exp}(\omega t + i k_x x + i k_y y)$, where $\omega$ is the growth rate, and $k_x, k_y$ are the wave numbers along the x- and y-axis. We find the critical strain for the onset of instability is (See {\color{blue}Supplemental Material} for detailed derivation)

\begin{equation}\label{eq:linearStablity}
    \epsilon > \frac{\snA_0}{B} \frac{\left[\snA_0 + K(\frac{2\pi}{L})^2 \right] \zeta}{\chi (\alpha - B)},
\end{equation}
indicating that the onset of instability is determined by a competition between an effective activity $\chi (\alpha -B)$ and the resistance of the passive material $\left[\snA_0 + K(\frac{2\pi}{L})^2 \right] \zeta$, similar to trends seen in isotropic solids subject to active stresses~\cite{yin2023contractility}. 
 
To corroborate our results we use numerical simulations of the full equations (\ref{eq:forceBalance},~\ref{eq:evolutionQ},~\ref{eq:evolutionA}) subject to periodic boundary conditions with initial random orientation and a homogeneously small nematic order $S_0$. The set of equations is implemented and solved using the finite element program FEniCS~\cite{fenics1,fenics2}. We observe a network-like pattern form over time, as is shown in Fig.~\ref{fig:2}(a). To understand how activity affects the formation of the network patterns, we perform a series of simulations at various cell contractions $\alpha-B$, treating the material as weakly compressible using the standard $u-p$ formulation (See {\color{blue}Supplemental Material} for implementation details). To capture the pattern, we plot the variance of the nematic order parameter $\sigma^2_S = \frac{1}{\Omega_0} \int_\Omega \left(S-\langle S \rangle\right)^2 d\Omega$, where $\langle S \rangle = \frac{1}{\Omega_0} \int_\Omega S d\Omega$ is the average of the nematic order parameter, and $\Omega_0=1$ is the area of the simulated unit domain. Notably, the patterns are inhibited at low activity, and the patterns form at moderate to high activity [Fig.~\ref{fig:2}(b)]. Furthermore, we plot $\sigma^2_S$ as a function of $\alpha-B$ at $\frac{t}{\tau}=300$, and see the appearance of a first-order phase transition around $\alpha-B=0.4$ can be observed [Fig.~\ref{fig:2}(c)]. We note that the first-order transition in nematic ordering arises even though the free energy $\Psi_{\mat}$ does not have a cubic term (as might have been expected), but occurs because of the non-trivial nematoelastic coupling. To verify this using numerical simulations, we plot $\sigma^2_S$ as a function of the stress-coupling coefficient $\chi$ and the externally applied strain $\epsilon$. The onset of instability is marked by a sharp transition in $\sigma^2_S$, which matches the linear stability theory (Fig.~\ref{fig:A1}).

\begin{figure*}[t]
    \centering
    \includegraphics[width=0.95\textwidth]{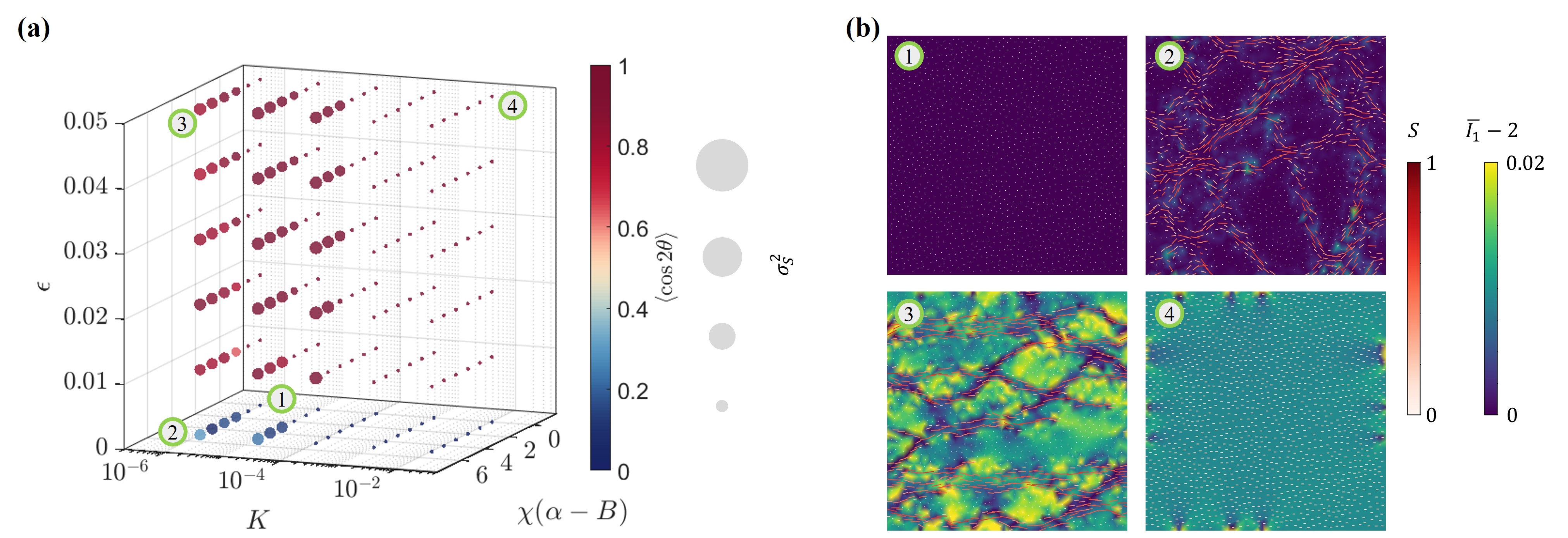}
    \caption{(a) Phase diagram. The color map indicates the degree of anisotropy $\langle \cos 2\theta \rangle$, and the marker size indicates the variance of nematic order parameter $\sigma^2_S$. 
    (b) The pattern becomes more anisotropic with larger $\epsilon$, highlighting the coupling between elasticity and nematicity. Representative steady-state configurations: 
    (1) without activity and without applied strain, $K=1\times 10^{-6}, \chi=0,\epsilon=0$, 
    (2) with activity and without applied strain, $K=1\times 10^{-6}, \chi=5,\epsilon=0$, 
    (3) with both activity and applied strain, $K=1\times 10^{-6}, \chi=5,\epsilon=0.05$, 
    (4) without activity and with applied strain, $K=1\times 10^{-2}, \chi=0,\epsilon=0.05$.
    The quivers indicate the director fields. 
    The color of the quiver indicates the magnitude of the nematic order parameter $S$, and the color of the background indicates the magnitude of the deformation $\bar{I}_1 = \text{tr}\, \bar{\C}$.
    The results are produced by numerically solving Eqs. (\ref{eq:forceBalance},~\ref{eq:evolutionQ},~\ref{eq:evolutionA}), and the rest of the parameters are $S_0=0.1, \snA_0 = 0.1, \frac{\gamma}{\tau}=1, \alpha-B=1, B = 0.5, \frac{t}{\tau}=100$.
    }
    \label{fig:4}
\end{figure*}

To extract a scaling law for the wavelength of the patterns seen in our simulations [Fig.~\ref{fig:2}(a)] as a consequence of the interactions between activity, nematic order, and elasticity, we consider a pattern with a characteristic size $\lambda$, and magnitude of perturbations in activity $\delta \snA$, nematic order $\delta S$, strain $\delta \epsilon$, and stress $\delta \sigma$. Eq.~\ref{eq:evolutionA} suggests that activity is promoted by the stress, i.e. $ \delta A \sim \delta \sigma$. From Eq.~\ref{eq:forceBalance} and Eq.~\ref{eq:stress}, we note that active contraction stresses balance the passive stresses due to deformation so that total stress is therefore given by $\delta \sigma = (\alpha-B) \delta S + \delta \epsilon = (\alpha-B-\alpha^{cr}) \delta S$. Finally, from Eq.~\ref{eq:evolutionQ}, we note that activity also controls nematic order, so that $S_0 \delta A \sim \frac{K}{\lambda^2} \delta S$. Combining these, we get
\begin{equation}\label{eq:powerLaw}
    \lambda \sim \left( \frac{K}{\alpha-B-\alpha^{cr}}\right)^{\frac{1}{2}}.
\end{equation}
Here $\alpha^{cr}>0$ is the critical effective activity for the onset of nematic patterns. Our numerical simulations for different values of $K$ with different active stress $\alpha-B$ confirm this scaling; the pattern size increases with increasing $K$, and decreases with increasing $\alpha-B$ ({\color{blue}Supplemental Material} Fig.S1). Furthermore, our numerical simulations are in good agreement with the pattern size given by the scaling relation in Eq.~\ref{eq:powerLaw} (Fig.~\ref{fig:3}). Finally, to elucidate pattern formation as a result of the coupling between activity, nematicity, and elasticity, we perform numerical simulation of the full nonlinear problem at a range of activity $\chi (\alpha - B)$, Frank constant $K$, and loading strain $\epsilon$. The full phase diagram is shown in Fig.~\ref{fig:4}.



\bigskip
\textit{Discussion.}---
Our theoretical and numerical framework indicates that cell activity (embodied as mechanical stresses) can give rise to the formation of network-like structures in cell-matrix mixtures, as a result of the coupling between activity, orientational ordering, and matrix elasticity. The theory sheds light on controlling nematic patterns in natural and synthetic multicellular living systems, and identifies potential biophysical therapeutic candidates to inhibit orientational ordering in diseased tissues, e.g. by varying the stiffness of the ECM, as well as the contractile activity of cells, qualitatively consistent with observations.  Our theory and numerical implementation set the stage to further explore a broad range of topics, such as how the activity-induced isotropic-nematic phase transitions might affect the bulk rheological properties of the whole cell-matrix mixture, and optimal control of functional active nematic elastomers. From an experimental perspective, it will be interesting to test whether the nematic patterns can be switched off by inhibiting cellular activity through treatments such as cytochalasin or blebbistatin, as well as whether removing the inhibition can recover the nematic patterns, problems for the future.

\section*{Acknowledgments}
We thank Sifan Yin and Yu Zhou for helpful discussions. H.Y. acknowledges support from MIT School of Engineering Takeda fellowship, M.G. acknowledges support from NIH (1R01GM140108), L.M. thanks the NSF-MRSEC 201175, the Simons Foundation and the Henri Seydoux Fund for partial financial support.

\section*{Code availability}
Our code will be available on a GitHub repository and a Google Colab notebook.

\appendix
\renewcommand\thefigure{A\arabic{figure}}   
\renewcommand{\thetable}{A\arabic{table}}
\setcounter{figure}{0}
\setcounter{table}{0}

\section{Stability analysis}

\begin{figure}[h]
    \centering
    \includegraphics[width=0.32\textwidth]{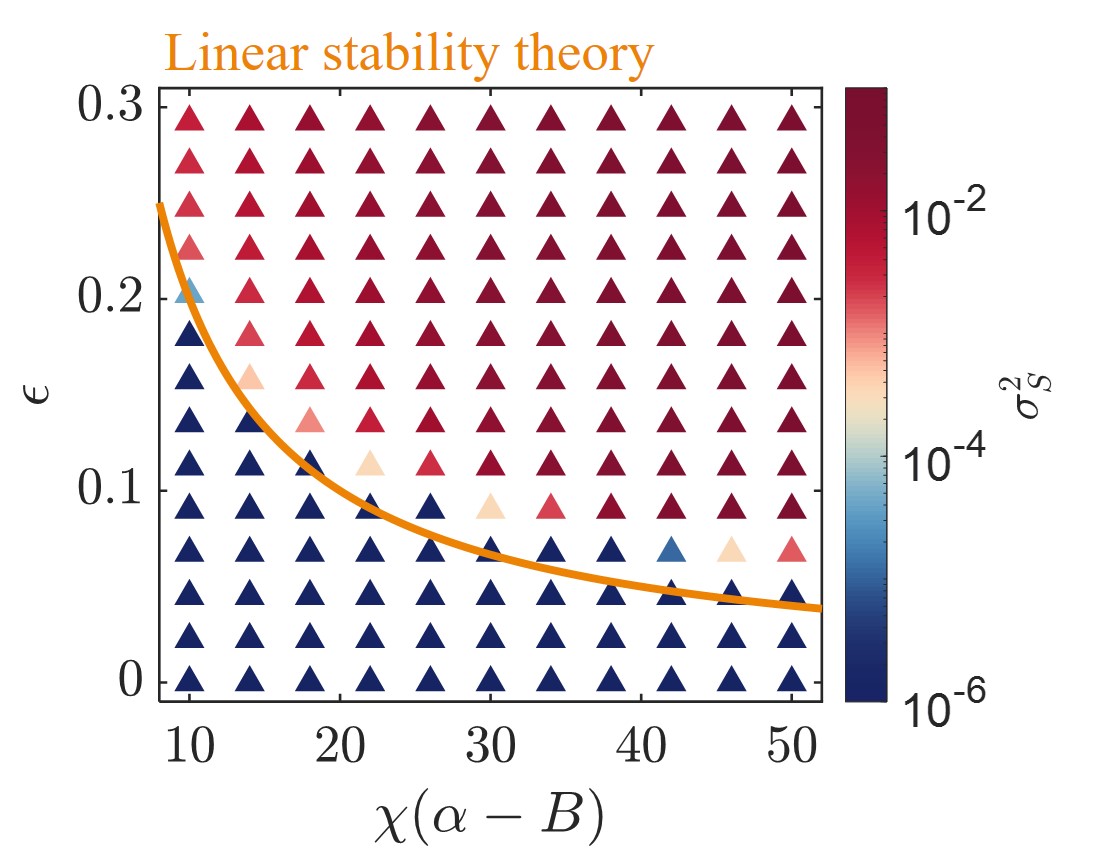}
    \caption{Stability under quasi-static external deformation $\epsilon$ at activity levels $\chi (\alpha - B)$. The orange line indicates the phase boundary given by linear stability analysis of a linearized incompressible version of the model (Eq.~\ref{eq:linearStablity}, details in the {\color{blue}Supplemental Material}).
    Here $S_0=0, \snA_0 = 1, \frac{\gamma}{\tau}=1, \alpha-B=1, B = 0.5, K=1 \times 10^{-4}$. 
    The numerical results are from solving the linearized form of Eqs. (\ref{eq:forceBalance},~\ref{eq:evolutionQ},~\ref{eq:evolutionA}) (Details in the {\color{blue}Supplemental Material}).
    }
    \label{fig:A1}
\end{figure}

The stability boundary with respect to changing activity $\chi (\alpha-B)$ and external strain $\epsilon$ is shown in Fig.~\ref{fig:A1}.

%
\bibliographystyle{unsrt}
\bibliography{biblio}

\end{document}